\documentclass{article}
  \usepackage{color}
\usepackage[dvips]{graphicx}
\usepackage{subfigure}

\def\no{\nonumber}
\def\lb{\label}
\def\a{{\alpha }}
\def\b{{\beta }}
\def\z{{\zeta }}
\def\be{\begin{equation}}
\def\ee#1{\label{#1}\end{equation}}

 \def\bx{\mathbf{x} }

 \def\bp{\mathbf{p} }
 
\def\no{\nonumber}
\def\lb{\label}

\def\no{\nonumber}
\def\lb{\label}

\def\q{\textsf{q} }

\newcommand{\ben}{\begin{eqnarray}}
\newcommand{\een}{\end{eqnarray}}
\date{}
\begin{document}

\title{THEORY AND APPLICATIONS OF THE RELATIVISTIC BOLTZMANN EQUATION}

\author{GILBERTO M. KREMER\footnote{kremer@fisica.ufpr.br}\\
Departamento de F\'{\i}sica, Universidade Federal do Paran\'a\\ Caixa Postal 19.044,
81.531-980 Curitiba,  Brazil
}

\maketitle

\begin{abstract}
In this work two systems are analyzed within the framework of the relativistic Boltzmann equation.  One of them refers to a description of binary mixtures of electrons and protons and of electrons and photons subjected to external electromagnetic fields in special relativity. In this case the Fourier and Ohm laws are derived and the corresponding transport coefficients are obtained. In the other  a relativistic gas under the influence of the Schwarzschild metric is studied. It is shown that the heat flux in Fourier's law in the presence of gravitational fields has three contributions, the usual dependence on the temperature gradient, and two relativistic contributions, one of them associated with an acceleration and another to a gravitational potential gradient. Furthermore it is shown that the transport coefficient of thermal conductivity decreases in the presence of a gravitational field. The dependence of the temperature field in the presence of a gravitational potential is also discussed.

\end{abstract}

\section{Introduction}	

The determination of the laws of Navier-Stokes, Fourier, Fick and Ohm from a non-relativistic kinetic theory based on the Boltzmann equation is an old subject in the literature dating back the begin of the last century. The relativistic version of the Navier-Stokes and Fourier laws from the Boltzmann equation were obtained later by Israel \cite{I} and Kelly \cite{Kl} in the sixties of last century. Relativistic gases are important to describe several astrophysical and cosmological problems like white dwarfs and neutron stars, particle production in the early universe, galaxy formation, controlled thermonuclear fusion, etc.

The aim of this work is to present the Boltzmann equations which describe  a gas under the influence of a gravitational field  and  a mixture of gases in the absence of gravitational fields but subjected to electromagnetic fields. Two system are analyzed, the first one consists in  binary mixtures of electrons and protons and of electrons and photons
subjected to external electromagnetic fields where the Fourier and Ohm laws are derived. The second one refers to the determination of Fourier's law for a relativistic gas under the influence of a  spherically symmetrical non-rotating and uncharged source of the gravitational field described by the Schwarzschild metric.

The work is structured as follows. In Section 2 the Boltzmann equations are introduced and in Section 3  binary mixtures of electrons and protons and of electrons and photons
subjected to external electromagnetic fields are analyzed. The relativistic gas under the influence the Schwarzschild metric is the subject of Section 4, and in the last section the main conclusions of this work are stated.

\section{Boltzmann equations}

As in the non-relativistic case, the relativistic Boltzmann equation describes the evolution of the one-particle distribution function of an ideal gas in the phase space spanned by the space-time coordinates $(x^\mu)=(x^0=ct, {\bf x})$ and momentum four-vectors $(p^\mu)=(p^0, {\bf p})$ of the particles. Due to the mass-shell condition $$p_0=\sqrt{g_{00}m^2c^2+\left(g_{0i}g_{0j}-g_{00}g_{ij}\right)p^ip^j}\qquad \hbox{and}\qquad p^0=(p_0-g_{0i}p^i)/ {g_{00}},$$ where $g_{\mu\nu}$ are the components of the metric tensor,
the one-particle distribution function depends only on $(\bx,\bp,t)$ and is defined in such a way that $f(\bx,\bp,t)d^3xd^3p$ gives at time $t$ the number of particles in the volume element $d^3xd^3p$.

For a single gas the Boltzmann equation reads (see e.g. \cite{CK})
\be
p^\mu\frac{\partial f}{\partial x^\mu}-\Gamma_{\mu\nu}^ip^\mu p^\nu
\frac{\partial f}{\partial p^i}=\mathcal{Q},
\ee{1}
where $\mathcal{Q}$ denotes the collision term, which depends on the product of the distribution functions of two particles at collision. Furthermore, $\Gamma_{\mu\nu}^i$ are Christofell symbols, since we are interested in  gases in the presence of gravitational fields. In the absence of it the Christofell symbols vanishes and (\ref{1}) reduces to the Boltzmann equation in a Minkowski space (special relativity).

For a mixture of ideal gases of $r$ constituents in the absence of gravitational fields but subjected to electromagnetic fields, the Boltzmannn equation for the $a-$component reads (see e.g. \cite{CK})
\ben\label{2}
p_a^{\mu}{\partial f_a\over \partial x^\mu}
+\frac{\q_a}{c}F^{\mu\nu}p_{a\nu}{\partial f_a\over \partial p_a^\mu}=\mathcal{Q}_a,\qquad a=1,\dots, r
\een
where $f_a\equiv f(\bx,\bp_a,t)$ denotes the one-particle distribution function of the $a-$component, $\q_a$ the particle's charge and $F^{\mu\nu}$ the electromagnetic field tensor. The collision term $\mathcal{Q}_a$ represents the collisions of the $a-$particles with all particles of the other constituents.

For simplicity we shall use in this work  model equations for the Boltzmann equations, which replace the collision terms by simpler expressions. In kinetic theory of relativistic gases two model equations are used, one of them was proposed  by Marle \cite{M1,M2} and another by Anderson and Witting \cite{AW}.

\section{Relativistic ionized gases}
In this section we follow \cite{KP} and  analyze  binary mixtures of electrons and protons and of electrons and photons
subjected to external electromagnetic fields within the framework of Anderson and Witting model equation.
These two systems are important in astrophysics since they could describe magnetic white dwarfs or
cosmological fluids in the plasma period and in the radiation dominated period.

\subsection{Basic fields}

In kinetic theory the macroscopic description of gas mixtures can be  represented the two first moments of the distribution function, namely, the partial particle four-flow $N_a^\alpha$ and the partial energy-momentum tensor $T_a^{\alpha\beta}$, which are defined by
\ben\label{3a}
&&N_a^\mu=c\int p_a^\mu f_a {d^3 p_a\over p_{a0}},
\\\label{3b} &&T_a^{\mu\nu}=c\int
p_a^\mu p_a^\nu f_a
{d^3 p_a\over p_{a0}}.
\een
The particle four-flow and the energy-momentum tensor of the mixture are given by the sums $N^\mu=\sum_a N_a^\mu$ and $T^{\mu\nu}=\sum_a T^{\mu\nu}_a$.

It is usual to introduce the four-velocity $U^\mu$ (such that $U^\mu U_\mu=c^2$) and the projector $\Delta^{\mu\nu}=g^{\mu\nu}-\frac{1}{c^2}U^\mu U^\nu$
and decompose the particle four-flows  and the  energy-momentum tensors  as
\ben\label{4a}
N_a^\mu=n_a U^\mu+J_a^\mu-\frac{n_a q^\mu}{nh},\qquad N^\mu=n U^\mu-\frac{q^\mu}{h},
\\\label{5a}
T_a^{\mu\nu}=-p_a\Delta^{\mu\nu}+{e_an_a\over c^2}U^\mu U^\nu+{1\over c^2}
U^\mu\left(q_a^\nu+h_a J_a^\nu-\frac{n_ah_a}{nh}q^\nu\right)
+{1\over c^2}
U^\nu\left(q_a^\mu+h_a J_a^\mu-\frac{n_ah_a}{nh}q^\mu\right),\\\label{5b} T^{\mu\nu}=-p\Delta^{\mu\nu}+{en\over c^2}U^\mu U^\nu.
\een
In the above decompositions -- known as  Landau and Lifshitz  decomposition \cite{LL} -- we have not taken into account two terms that appear in the  energy-momentum tensors (\ref{5a}) and (\ref{5b}), namely, the traceless part of the viscous pressure tensor and the non-equilibrium pressure, since we are only interested in deriving the laws of Fourier and Ohm.
Above we have introduced the following  quantities for the constituent $a$ in the mixture:  particle number density $n_a$,
diffusion flux  $J^\alpha_a$,  hydrostatic pressure
$p_a$,  heat flux $q_a^\alpha$,   energy per particle $e_a$ and
enthalpy per particle $h_a=e_a+p_a/n_a$. The corresponding quantities for the mixture are given by the sums
\ben\label{6}
n=\sum_{a=1}^r n_a,\qquad p=\sum_{a=1}^r p_a,\qquad ne=\sum_{a=1}^r n_ae_a,
\\\label{7}
q^\mu=\sum_{a=1}^r(q_a^\mu+h_aJ_a^\mu),\qquad n h=\sum_{a=1}^r n_a h_a.
\een
Furthermore, the diffusion fluxes are constraint by the sum $\sum_a J_a^\mu=0$.

The electromagnetic field tensor  is decomposed as (see e.g \cite{vv})
\be
F^{\mu\nu}={1\over c}\left(E^\mu U^\nu-
E^\nu U^\mu \right)-B^{\mu\nu},
\ee{8}
where $E^\mu$ is identified with the electrical field and $B^{\mu\nu}$ with the magnetic flux induction. They are given by the following projections of the electromagnetic field tensor
\be
E^\mu={1\over c}F^{\mu\nu}U_{\nu},\qquad
B^{\mu\nu}=-\Delta^\mu_{\sigma} F^{\sigma\tau}
\Delta^\nu_{\tau},
\ee{9}

The electric current four-vector
$I^\mu$  is defined in terms of the partial diffusion
fluxes $J_a^\mu$ and of the partial electric
charges $\q_a$ as
\be
I^\mu=\sum_{a=1}^r \q_a J_a^\mu.
\ee{10}

\subsection{The non-equilibrium distribution function}

For the derivation of the laws of
Fourier and Ohm for a binary mixture of electrons and protons and of
electrons and photons,  some simplifications of the model must be made, namely,
\begin{itemize}
\item [(a)] for a mixture of protons $(a=p)$ and electrons $(a=e)$, the ratio of the rest masses is $m_p/m_e\approx 1836$.  This condition defines a Lorentzian plasma (see e. g. \cite{Lif}), where
the collisions between  the electrons may be neglected in comparison with the collisions between the electrons and protons. The electric charges are given
by $\q_e=-\q_p=-{\rm e}$,  where ${\rm e}$ denotes the elementary charge, and if we consider a locally neutral system where
$\q_en_e+\q_pn_p=0$ we have  that $n_e=n_p$. Due to the relationship between the diffusion fluxes
$J_e^\mu=-J_p^\mu$, the electric current four-vector (\ref{10}) reads
\be
I^\mu=-2{\rm e}J^\mu_e;
\ee{11}

\item [(b)] for a binary mixture of electrons $(a=e)$ and
photons $(a=\gamma)$ the collisions between electrons can also be neglected in
comparison to the the Compton scattering, which refers to the collisions between electrons and photons. In this case
the electric current four-vector
(\ref{10}), reduces to
\be
I^\mu=-{\rm e}J^\mu_e,
\ee{12}
since the electric charge of the
photons is zero $(\q_\gamma=0)$;

\item [(c)] the heat flux of the mixture (\ref{9})$_1$ reduces to
\be
q^\mu=q_e^\mu+(h_e-h_p)J^\mu_e,\qquad \hbox{or} \qquad q^\mu=q_e^\mu+(h_e-h_\gamma)J^\mu_e,
\ee{13}
due to the fact that the partial heat fluxes of the protons and of the photons are negligible
in comparison with the partial heat flux of the electrons.
\end{itemize}

The Anderson and Witting model of the Boltzmann equation (\ref{2})  for electrons that follow the above simplifications reads
\be
p_e^\mu {\partial f_e\over \partial x^\mu}
-{{\rm e}\over c}F^{\mu\nu}p_{e\nu}{\partial f_e \over \partial p_e^\mu}
=-{U^\sigma p_{e\sigma}\over c^2\tau_{eb}}(f_e-f_e^{(0)}),\qquad b=p,\gamma.
\ee{14}
Here $\tau_{ep}$ and  $\tau_{e\gamma}$ are the mean free time between collisions of
electrons-protons or electrons-photons, respectively, and
$f_e^{(0)}$ is the equilibrium distribution function for the electrons. The electrons obey the Fermi-Dirac statistics so that the equilibrium distribution function reads (see e.g. \cite{CK})
\be
f_e^{(0)}={2\over h^3}{1\over \exp\left({-{\mu_e\over kT}
+{U^\alpha p_{e\alpha}\over kT}}\right)+ 1},
\ee{15}
where $h$ denotes Planck's constant, $k$ Boltzmann's constant, $T$ the temperature of the mixture, $\mu_e$  the chemical potential of the electrons and the factor 2
refers to the degeneracy factor of the electrons.

From the knowledge of  the equilibrium distribution function of the electrons it is possible to determine the values of particle number density $n_e$, energy per particle $e_e$ and pressure  $p_e$ of the electrons, yielding
\ben\label{16}
n_e&=&\frac{1}{c^2} U_\a N_e^\a=\frac{1}{c^2} U_\a\, c\int p_e^\a f_e^{(0)}\frac{d^3 p_e}{p_{e0}}=\frac{8\pi}{h^3}\left(m_ec\right)^3\mathcal{J}_{21}(\z_e,\mu_e^\star),\\
e_e&=&\frac{1}{n_ec^2} U_\a U_\b T_e^{\a\b}=\frac{1}{n_ec^2} U_\a U_\b \,c\int p_e^\a p_e^\b f_e^{(0)}\frac{d^3 p_e}{p_{e0}}=\frac{8\pi}{n_eh^3}m_e^4c^5\mathcal{J}_{22}(\z_e,\mu_e^\star),\\
p_e&=&-\frac{1}{3} \Delta_{\a\b} T_e^{\a\b}=-\frac{1}{3} \Delta_{\a\b} \,c\int p_e^\a p_e^\b f_e^{(0)}\frac{d^3 p_e}{p_{e0}}=\frac{8\pi}{h^3}m_e^4c^5\mathcal{J}_{40}(\z_e,\mu_e^\star).
\een
In the above equations  $\mathcal{J}_{nm}(\z_e,\mu_e^\star)$ denotes to the integral
\ben\label{17}
\mathcal{J}_{nm}(\z_e,\mu_e^\star)=\int_0^\infty \frac{\sinh^n \vartheta\cosh^m\vartheta d\vartheta}{\exp(-\mu_e^\star+\z_e\cosh\vartheta)+1},
\een
where $\mu_e^\star=\mu_e/kT$ represents the electron chemical potential in units of $kT$ and $\z_e=m_ec^2/kT$ is the ratio between the electron rest  energy $m_ec^2$ and the thermal energy of the gas $kT$. For $\z_e\gg1$ the electrons behave as  non-relativistic, and for $\z_e\ll1$ they behave as ultra-relativistic.

For the determination of the
non-equilibrium distribution function for the electrons from the Boltzmann equation (\ref{14}) we adopt the Chapman-Enskog method and
search for a solution  of the form $f_e=f_e^{(0)}+\phi_e$,
where the deviation from the equilibrium distribution function is
considered to be a small quantity, i.e., $\vert\phi_e\vert<1$. By inserting the representation $f_e=f_e^{(0)}+\phi_e$ into
 (\ref{14}) and performing the derivatives we get
 \ben\nonumber
{-2\over h^3}{\exp\left({-{\mu_e\over kT}
+{U^\alpha p_{e\alpha}\over kT}}\right)\over \left[\exp\left({-{\mu_e\over kT}
+{U^\alpha p_{e\alpha}\over kT}}\right)+ 1\right]^2}\left\{{1\over c^2}
(p_e^\nu U_\nu)\left[D\left({\mu_e\over kT}\right)+{p_e^\mu
U_\mu\over kT^2}DT\right]
+{p_e^\nu U_\nu\over kT^2}p_{e\mu}\left[\nabla^\mu T
-{T\over c^2}DU^\mu\right]\right.\\\label{18}
\left.-{{\rm e}\over kT}p_{e\mu}\left[E^\mu
 -{kT\over {\rm e}}
\nabla^\mu \left({\mu_e\over kT}\right)
\right]
-{{p_{e\mu} p_{e\nu}\over kT}\nabla^\mu U^\nu}
\right\}
={U^\gamma p_{e\gamma}\over c^2\tau_{eb}}\left[ 1+{m_ec\over
U^\delta p_{e\delta}}\left({\omega_e\tau_{eb}\over B}\right)B^{\mu\nu}
p_{e\nu}{\partial\over \partial p_e^\mu}\right]\phi_e.
\een
Here we have not taken into account the terms ${\partial \phi_e/ \partial x^\alpha}$ and $E^\alpha{\partial\phi_e/\partial
p_e^\alpha},$ since  our aim  is to derive linear constitutive equations with first order gradients and electric field.
Moreover, it was introduced the differential operators $D\equiv U^\alpha\partial_\alpha$ and
$\nabla^\alpha\equiv\Delta^{\alpha\beta}\partial_\beta$ and  the electron cyclotron frequency
$\omega_e={\rm e}B/m_e$, with $B$ denoting the modulus of the magnetic flux induction.

Since the aim of this work is to derive  Fourier and Ohm laws, we fix our attention to the  thermodynamic forces that appear in the deviation
from the equilibrium distribution function which are the four-vectors:
\ben\label{19a}
&&\nabla^\mu{\cal T}\equiv \left[\nabla^\mu T-{T\over c^2}DU^\mu\right],
\\\label{19b}
&&{\cal E}^\mu\equiv \left[E^\mu-{kT\over {\rm e}}
\nabla^\mu \left({\mu_e\over kT}\right)
\right].
\een
The first one is a combination of a temperature gradient and an
acceleration term and the second one refers to a combination of an external electric
field and a gradient of the chemical potential of the electrons.
Hence, up to terms in $(\omega_e\tau_{eb}/B)^2$  we obtain from (\ref{18}) that the deviation from the distribution
function reads
\ben\nonumber
\phi_e={-2\over h^3}{\exp\left({-{\mu_e\over kT}
+{U^\alpha p_{e\alpha}\over kT}}\right)\over \left[\exp\left({-{\mu_e\over kT}
+{U^\alpha p_{e\alpha}\over kT}}\right)+ 1\right]^2}{c^2\tau_{eb}\over
U^\gamma p_{e\gamma}}\left[\eta^{\mu\nu}-{m_ec\over U^\delta
p_{e\delta}}\left({\omega_e\tau_{eb}\over B}\right)B^{\mu\nu}\right.
\\\label{20}
\left.+\left({m_ec\over U^\delta p_{e\delta}}\right)^2\left({\omega_e\tau_{eb}
\over B}\right)^2B^{\mu}_\sigma B^{\sigma\nu}\right]p_{e\nu}\left\{{p_e^\beta U_\beta\over kT^2}\nabla_\mu{\cal T}
-{{\rm e}\over kT}{\cal E}_\mu\right\}.
\een

Once we know the deviation from the equilibrium distribution function (\ref{20}) we proceed to obtain the laws of Ohm and Fourier.

\subsection{Fourier and Ohm laws}

The diffusion flux $J_e^\mu$ and the heat flux
$q_e^\mu$ of the electrons follows from the projections of (\ref{3a}) and (\ref{3b}) by using the decompositions (\ref{4a}) -- (\ref{5b}), namely,
\be
{h_b\over h}J_e^\mu-{n_e\over nh}q_e^\mu=\Delta^\mu_\nu
N_e^\nu=\Delta^\mu_\nu \int c
p_e^\nu f_e{d^3p_e\over p_{e0}},
\ee{21}
\be
{h_bh_e\over h}J_e^\mu+{n_bh_b\over nh}q_e^\mu=
\Delta^\mu_\nu U_\gamma T_e^{\nu\gamma}=
\Delta^\mu_\nu U_\gamma\int c
p_e^\nu p_e^\gamma f_e{d^3p_e\over p_{e0}}.
\ee{22}
If we  insert the distribution function of the electrons in (\ref{21}) and (\ref{22}) and integrate the
resulting equations it follows a system of equations for $J_e^\mu$ and
$q_e^\mu$ which is used to determine the heat flux of the mixture
(\ref{13}) and the electric current four-vector of the mixture of electrons and protons (\ref{11}) or of the mixture of electrons and photons
(\ref{12}). The laws of  Fourier and Ohm are written respectively as
\ben\label{23a}
q^\mu=\Lambda^{\mu\nu}\nabla_\nu {\cal T}+
\Upsilon^{\mu\nu}{\cal E}_\nu,
\\\label{23b}
I^\mu=\sigma^{\mu\nu}{\cal E}_\nu+\Omega^{\mu\nu}
\nabla_\nu {\cal T}.
\een
We may identify the tensors   $\Lambda^{\mu\nu}$ with the thermal conductivity, $\sigma^{\mu\nu}$ with the
electrical conductivity, while
 $\Upsilon^{\mu\nu}$ and $\Omega^{\mu\nu}$ with  cross effects. We may represent the general expressions for the above
mentioned tensors as
\ben\lb{24}
\left(
  \begin{array}{c}
  \Lambda^{\mu\nu} \\
  \Upsilon^{\mu\nu} \\
  \sigma^{\mu\nu} \\
  \Omega^{\mu\nu}
  \end{array}
\right)=
\left(
  \begin{array}{c}
    a_1 \\
    b_1 \\
    c_1 \\
    d_1 \\
  \end{array}
\right)\eta^{\mu\nu}+
\left(
  \begin{array}{c}
    a_2 \\
    b_2 \\
    c_2 \\
    d_2 \\
  \end{array}
\right)B^{\mu\nu}+
\left(
  \begin{array}{c}
    a_3 \\
    b_3 \\
    c_3 \\
    d_3 \\
  \end{array}
\right)
B^{\mu\gamma}B_\gamma^\nu,
\een
where the scalar coefficients $a_1$ through $d_3$ are expressed in terms of the integrals (\ref{17}).

The true thermal conductivity tensor $\lambda^{\mu\nu}$
is obtained  by assuming that there is no
electric current. In this case we eliminate ${\cal E}^\mu$ from (\ref{23a}) through the use of (\ref{23b}) and  write Fourier law as
\be
q^\mu=\lambda^{\mu\nu}\nabla_\nu{\cal T},\quad\hbox{where}\quad
\lambda^{\mu\nu}=\Lambda^{\mu\nu} -\Upsilon^{\mu\sigma}\left(\sigma^{-1}\right)_{\sigma\tau}\Omega^{\tau\nu}.
\ee{25}

It is usual in the theory of ionized gases
to decompose the thermodynamic forces $\nabla^\mu{\cal T}$
and ${\cal E}^\mu$ into parts
parallel $(\parallel)$, perpendicular $(\perp)$ and
transverse $(t)$ to the magnetic flux induction as follows
\ben\label{26}
\left(
  \begin{array}{c}
    \nabla^\mu_\parallel{\cal T} \\
    {\cal E}^\mu_\parallel \\
  \end{array}
\right)&=&
{\widetilde B^{\mu\nu}\widetilde B_{\nu\tau}\over \left({1\over 2}B^{\gamma\delta}
B_{\gamma\delta}\right)}
\left(
  \begin{array}{c}
    \nabla^\tau{\cal T} \\
    {\cal E}^\tau \\
  \end{array}
\right),\\\left(
  \begin{array}{c}
    \nabla^\mu_t{\cal T} \\
    {\cal E}^\mu_t \\
  \end{array}
\right)&=&
{B^{\mu\nu}\over \left({1\over 2}B^{\gamma\delta}
B_{\gamma\delta}\right)}
\left(
  \begin{array}{c}
    \nabla_\nu{\cal T} \\
    {\cal E}_\nu \\
  \end{array}
\right),
\\\label{27}
\left(
  \begin{array}{c}
    \nabla^\mu_\perp{\cal T} \\
    {\cal E}^\mu_\perp \\
  \end{array}
\right)&=&-
{1\over \left({1\over 2}B^{\gamma\delta}
B_{\gamma\delta}\right)}B^{\mu\nu} B_{\nu\tau}
\left(
  \begin{array}{c}
    \nabla^\tau{\cal T} \\
    {\cal E}^\tau \\
  \end{array}
\right),
\een
where $\widetilde B^{\mu\nu}=\epsilon^{\mu\nu\sigma\tau}B_{\sigma\tau}/2$ is the dual of the magnetic flux induction tensor $B^{\mu\nu}$. In a local Lorentz rest frame these decompositions become
\ben\label{28}
&&\left(
  \begin{array}{c}
    \nabla^0_\parallel{\cal T} \\
    {\cal E}^0_\parallel \\
  \end{array}
\right)
=\left(
  \begin{array}{c}
    \nabla^0_\perp{\cal T} \\
    {\cal E}^0_\perp \\
  \end{array}
\right)
=\left(
  \begin{array}{c}
    \nabla^0_t{\cal T} \\
    {\cal E}^0_t \\
  \end{array}
\right)
=\left(
  \begin{array}{c}
    0 \\
    0 \\
  \end{array}
\right),\\&& \left(
  \begin{array}{c}
    \nabla^i_\parallel{\cal T} \\
    {\cal E}^i_\parallel \\
  \end{array}
\right)
=\frac{B_j}{B^2}\left(
  \begin{array}{c}
    \nabla^j{\cal T} \\
    {\cal E}^j \\
  \end{array}
\right)B^i,
\\&&
\left(
  \begin{array}{c}
    \nabla^i_\perp{\cal T} \\
    {\cal E}^i_\perp \\
  \end{array}
\right)
=\frac{B_j}{B^2}\left(
  \begin{array}{c}
    \nabla^j{\cal T} \\
    {\cal E}^j \\
  \end{array}
\right)B^i-\left(
  \begin{array}{c}
    \nabla^i{\cal T} \\
    {\cal E}^i \\
  \end{array}
\right),
\\&&
\left(
  \begin{array}{c}
    \nabla^i_t{\cal T} \\
    {\cal E}^i_t \\
  \end{array}
\right)
=\frac{\epsilon^{ijk}B_k}{B}\left(
  \begin{array}{c}
    \nabla_j{\cal T} \\
    {\cal E}_j \\
  \end{array}
\right).
\een
From the above equations it is easy to verify that $\nabla^i_\parallel{\cal T},
    {\cal E}^i_\parallel$ are parallel to the magnetic flux induction $B^i$,
$\nabla^i_\perp{\cal T}, {\cal E}^i_\perp$ are perpendicular to it, while $\nabla^i_t{\cal T},
{\cal E}^i_t$ are perpendicular to the parallel and perpendicular decompositions.

In terms of these decompositions  the electric current four-vector and the heat flux can be written, without the cross-effects terms,  respectively as
\ben
I^\mu&=&\sigma_\parallel {\cal E}^\mu_\parallel+\sigma_\perp
{\cal E}^\mu_\perp+\sigma_t {\cal E}^\mu_t,\\
q^\mu&=&\lambda_\parallel
\nabla^\mu_\parallel{\cal T}+\lambda_\perp
\nabla^\mu_\perp{\cal T}+\lambda_t\nabla^\mu_t{\cal T}.
\een
Here the scalars  are called the
parallel, perpendicular and transverse components of the  tensors of thermal and electrical conductivity.

\subsection{Thermal and electrical conductivities}

\subsubsection{Non-degenerate electrons}

Here we shall analyze two  cases: (a)   a non-relativistic mixture of protons and non-degenerate electrons and (b) an ultra-relativistic mixture of non-degenerate electrons and photons. For non-degenerate electrons the chemical potential must fulfill the condition that  $e^{-\mu_e^\star}\gg1$.
\begin{itemize}
\item [(a)] Two conditions define a non-relativistic mixture of electrons and protons, namely, $m_p/m_e\gg1$ and $\zeta_e=m_ec^2/(kT)\gg1$. For this case the
transport coefficients  become
\ben\label{31}
&&\sigma_\parallel={e^2\tau_{ep}n_e\over m_e }\left(1-{5\over 2\zeta_e}
\right),\qquad \sigma_t=\sigma_\parallel
\left(1-{5\over 2\zeta_e}\right)(\omega_e\tau_{ep}),
\\\label{32}
&&\sigma_\perp=\sigma_\parallel\left[1
-\left(1-{5\over \zeta_e}\right)(\omega_e\tau_{ep})^2\right],\\
&&\lambda_\parallel={5k^2T\tau_{ep}n_en\over2 m_en_p }\left(1-{3\over \zeta_e}
\right),
\qquad
\lambda_t=\lambda_\parallel
\left(1-{9\over 2\zeta_e}\right)(\omega_e\tau_{ep}),
\\
&&\lambda_\perp=\lambda_\parallel\left[1-\left(1-{9\over \zeta_e}\right)(\omega_e\tau_{ep})^2\right].
\een
Without the relativistic
corrections given by $1/\zeta_e$, the thermal and electrical conductivities reduce to the  coefficients
of non-degenerate and non-relativistic ionized gases, namely,
\ben\label{34}
\sigma_\parallel={e^2\tau_{ep}n_e\over m_e },\qquad \sigma_t=\sigma_\parallel
(\omega_e\tau_{ep}),
\qquad
\sigma_\perp=\frac{\sigma_\parallel}{1+(\omega_e\tau_{ep})^2},
\\\label{35}
\lambda_\parallel={5k^2T\tau_{ep}n_en\over2 m_en_p },
\qquad
\lambda_t=\lambda_\parallel(\omega_e\tau_{ep}),
\qquad
\lambda_\perp=\frac{\lambda_\parallel}{1+(\omega_e\tau_{ep})^2},
\een
by considering the condition $\omega_e\tau_{ep}<1$.

\item [(b)] An ultra-relativistic mixture of  non-degenerate electrons and photons is characterized by the condition  $\zeta_e=m_ec^2/(kT)\ll1$. Here  the
transport coefficients reduce to
\ben\label{36}
\sigma_\parallel=\sigma_\perp={e^2c^2\tau_{e\gamma}n_e(3n_e+4n_\gamma)\over
12nkT},
\quad \sigma_t=\sigma_\parallel{n_e+2n_\gamma\over 3n_e+4n_\gamma}(\omega_e\tau_{e\gamma})
\zeta_e,
\\\label{37}
\lambda_\parallel=\lambda_\perp={4kc^2\tau_{e\gamma}n_en\over 3n_e+4n_\gamma},
\qquad \lambda_t=\lambda_\parallel{2n
\over (3n_e+4n_\gamma)}(\omega_e\tau_{e\gamma})\zeta_e.
\een
Hence the parallel and perpendicular electrical and thermal conductivities
coincide, while the transverse electrical and thermal conductivities are small quantities since they
are proportional to $\zeta_e$.
\end{itemize}

\subsubsection{Completely degenerate electrons}

In the limiting case of completely degenerate electrons all thermal conductivities vanish. This behavior is connected with the well-known result
from statistical mechanics that the heat capacity  of a completely
degenerate gas vanishes. There exist  three  cases to be analyzed  which
are: (a) a  non-relativistic mixture of protons and completely degenerate
electrons; (b) an ultra-relativistic mixture of completely
degenerate  electrons and photons and (c) a mixture of non-relativistic protons and
ultra-relativistic completely degenerate  electrons. We proceed to analyze
the electrical conductivities for these cases.
\begin{itemize}

\item [(a)] A  non-relativistic mixture of protons and completely
degenerate electrons is identified by $\zeta_e\gg1$ and $p_F\ll m_ec$, where
$p_F$ denotes the Fermi momentum of the electrons. Here we have
\be
\sigma_\parallel={8\pi e^2\tau_{ep}p_F^3\over 3m_e h^3},\qquad
\sigma_t=\sigma_\parallel(\omega_e\tau_{ep}),
\qquad
\sigma_\perp={\sigma_\parallel\over 1+(\omega_e\tau_{ep})^2},
\ee{38}
which  show the dependence of the transverse and perpendicular electrical conductivities on the magnetic flux induction $B$ through
the electron cyclotron frequency $\omega_e$.

\item [(b)] An ultra-relativistic mixture of  completely degenerate  electrons and photons
is characterized by the conditions $\zeta_e\ll1$ and $p_F\gg m_ec$, and the electrical conductivities for this case read
\ben\label{39}
\sigma_\parallel=\sigma_\perp={8\pi e^2\tau_{e\gamma}c^2n_ep_F^3\over
12nkTh^3 }\left(1+{4kTn_\gamma\over n_ecp_F}\right),
\qquad
\sigma_t=\sigma_\parallel\frac{m_ec}{p_F}(\omega_e\tau_{e\gamma}).
\een
Here the parallel and perpendicular
electrical conductivities are equal to each other, while the transverse electrical
conductivity is a small quantity since it is proportional to $m_ec/p_F$.

\item [(c)] A mixture of non-relativistic protons and ultra-relativistic
completely degenerate  electrons is also an important case, since
it could describe a white dwarf star. Here  the conditions
$m_p/m_e\gg1$ and $p_F\gg m_ec$ hold and the electrical conductivities become
\be
\sigma_\parallel=\sigma_\perp={8\pi e^2\tau_{ep} cp_F^2\over 3h^3 },\qquad
\sigma_t=\sigma_\parallel\frac{m_ec}{p_F}(\omega_e\tau_{e\gamma}),
\ee{40}
showing that the parallel and perpendicular conductivities coincide and that
the transverse conductivity is a small quantity, since it is proportional to $m_ec/p_F$.
\end{itemize}

\section{Relativistic gas in a gravitational field}

In this section we are interested in the analysis of a relativistic gas in the presence of a gravitational field. We follow \cite{K} and consider that the gas is not the source of the gravitational field but is under the influence of a Schwarzschild metric, which is the solution of Einstein's field equation for  a  spherically symmetrical non-rotating and uncharged source of the gravitational field.

\subsection{Basic equations}

Here we shall use the isotropic Schwarzschild metric which reads (see e.g. \cite{A})
\ben\lb{41}
 ds^2=g_0(r)\left(dx^0\right)^2-g_1(r)\delta_{ij}dx^idx^j.
 \een
Above we have introduced the following abbreviations
 \be
g_0(r)=\frac{\left(1+\frac{\Phi}{2c^2}\right)^2}{\left(1-\frac{\Phi}{2c^2}\right)^2},\qquad
g_1(r)=\left(1-\frac{\Phi}{2c^2}\right)^4,\qquad\hbox{where}\qquad \Phi=-\frac{GM}{r},
\ee{42}
is the gravitational potential, $G$  the gravitational constant and $M$  the mass of the spherical source.

The Marle model of the Boltzmann equation in the presence of gravitational fields (\ref{1}) reads
\be
p^\mu\frac{\partial f}{\partial x^\mu}-\Gamma_{\mu\nu}^ip^\mu p^\nu
\frac{\partial f}{\partial p^i}=-\frac{m}{\tau}\left(f-f^{ (0)}\right),
\ee{43}
where $\tau$ denotes a mean free time and $f^{(0)}$   the Maxwell-J\"uttner distribution function which in a comoving frame -- where $U^\mu=(c/\sqrt{g_0},\bf0)$ -- becomes
\be
f^{(0)}=\frac{n}{4\pi k T m^2 c  K_2(\z)}\exp\left({-\frac{c\sqrt{m^2c^2+g_1\vert\bp\vert^2}}{k T}}\right).
\ee{44}
Here $$ K_n(\zeta)=\left(\frac{\zeta}{2}\right)^n\frac{\Gamma(1/2)}{\Gamma(n+1/2)}\int_{1}^\infty e^{-\zeta y}\left(y^2-1\right)^{n-1/2}\,dy$$ is the modified Bessel function of second kind
and $\z=mc^2/kT$.

In Marle's model the particle four-flow and the energy-momentum tensor
defined in terms of the one-particle distribution function by
\ben
&&N^\mu= c\int p^\mu f\,\sqrt{-g}\,\frac{d^3p}{p_0},\\
&&T^{\mu\nu}= c\int p^\mu p^\nu f\,\sqrt{-g}\,\frac{d^3p}{p_0},
\een
 are written in terms of Eckart's decomposition \cite{Eck} as
\ben\lb{45}
&&N^{\mu}=nU^{\mu},\\
&&T^{\mu\nu}=\sigma^{\mu\nu}-\left(p+\varpi\right)
\Delta^{\mu\nu}
+\frac{1}{c^2}\left(q^\mu U^\nu+q^\nu U^\mu\right)+\frac{en}{c^2}U^{\mu} U^{\nu},
\een
where $\varpi$ denotes the non-equilibrium pressure and $\sigma^{\mu\nu}$ the traceless part of the viscous pressure tensor.

\subsection{The non-equilibrium distribution function}

In order to determine the non-equilibrium distribution function we rely on the Chapman-Enskog method and write the distribution function as $f=f^{(0)}(1+\varphi)$, where $\vert\varphi\vert<1$ is a small deviation from the Maxwell-J\"uttner distribution function. In the Chapman-Enskog method the equilibrium distribution function is inserted on the left-hand side of Boltzmann equation (\ref{43}) and the representation $f=f^{(0)}(1+\varphi)$ on its right-hand side. By performing the derivatives we obtain that the non-equilibrium distribution function is given by
\ben\no
-\frac{m}{\tau}\left(f-f^{ (0)}\right)=-\frac{m}{\tau}f^{ (0)}\varphi=f^{ (0)}\bigg\{\frac{p^\nu}{n}\frac{\partial n}{\partial x^\nu}
+\frac{p^\nu}{T}\left[1-\frac{K_3\z}{K_2}+\frac{ p^\tau U_\tau}{kT}\right]\frac{\partial T}{\partial x^\nu}
\\\lb{46}
-\frac{p^i p^\nu}{kT}\frac{\partial U_i}{\partial x^\nu}
-\frac{c^2}{2kT}\frac{d g_1}{dr}\frac{p^ip^jp^k}{U^\tau p_\tau}\delta_{ij}\delta_{kl}\frac{x^l}{r}
+\frac{c^2}{k T}g_1\delta_{ij}\Gamma_{\sigma\nu}^i \frac{p^j p^\sigma p^\nu }{U^\tau p_\tau}\bigg\}.
\een

\subsection{Field equations for a non-viscous and non-conducting gas}

From (\ref{46}) it is possible to obtain the field equations for the particle number density $n$, four-velocity $U^\mu$ and temperature $T$ of  a non-viscous and non-conducting relativistic gas in the presence of a gravitational field.

First we multiply  (\ref{46}) by $\sqrt{-g}d^3 p/p_0$, integrate the resulting equation and obtain the particle number density balance equation:
\be
U^\nu\,\frac{\partial n}{\partial x^\nu}+n\, {U^\nu}_{;\nu}=0.
\ee{47}

If we multiply (\ref{46}) by $p^\mu\sqrt{-g}d^3 p/p_0$ and integrate the resulting equation it follows:
\begin{enumerate}
\item the momentum density balance equation from the projection $\Delta^\nu_\mu$:
\ben\lb{48}
m\,n\,\frac{K_3}{K_2}U^\mu \frac{\partial U_i}{\partial x^\mu}-\frac{\partial p}{\partial x^i}-m\,n\,\frac{K_3}{K_2}\frac{1}{1-\Phi^2/4c^4}\frac{\partial\Phi}{\partial x^i}=0,
\een
where the first term refers to an acceleration, the second one to a pressure gradient and the third one to a gravitational potential gradient;
\item the energy density balance equation from the projection $U_\mu$:
\ben\label{49}
n \,c_v\,U^\nu\,\frac{\partial T}{\partial x^\nu}+p\, {U^\nu}_{;\nu}=0,
\een
where the heat capacity per particle at constant volume is given by
\ben\lb{50}
c_v=k\left(\zeta^2+5\frac{K_3}{K_2}\zeta-\frac{K_3^2}{K_2^2}\zeta^2-1\right).
\een
\end{enumerate}

In the non-relativistic limiting case ($\z\gg1$) and in the presence of a weak gravitational field ($\Phi/c^2\ll1)$, eq. (\ref{47}) reduces to Newton's second law in the presence of a gravitational field:
\ben\lb{51}
m\,n\,U^\mu \frac{\partial U_i}{\partial x^\mu}-\frac{\partial p}{\partial x^i}-m\,n\,\frac{\partial\Phi}{\partial x^i}=0.
\een

Equations (\ref{46}) -- (\ref{48}) are the field equations of an Eulerian relativistic gas -- i.e.,  a non-viscous and non-conducting relativistic gas --  described by the basic fields
$(n, U^\mu, T)$.

\subsection{Fourier's law}

Let us analyze Fourier's law in the presence of a gravitational field. If we insert the distribution function, determined in the last section into the definition of the energy-momentum tensor we get
\ben\lb{52}
q^i=\Delta_\mu^i\, U_\nu\,T^{\mu\nu}=\Delta_\mu^i\, U_\nu\,\int \,c\,p^\mu p^\nu f^{(0)}(1+\varphi)\sqrt{-g}\frac{d^3p}{p_0}.
\een
By using (\ref{46}) and performing the integration it follows Fourier's law
\ben\lb{53}
q^i=-\lambda \delta^{ij}\left[\frac{\partial T}{\partial x^j}-\frac{T}{c^2}\left(U^\sigma \frac{\partial U_j}{\partial x^\sigma}+\frac{1}{1-\Phi^2/4c^4}\frac{\partial\Phi}{\partial x^j}\right)\right],
\een
where $\lambda$ --  the coefficient of thermal conductivity -- is given by
\ben\lb{54}
\lambda=\frac{\tau p}{\left(1+\frac{\vert\Phi\vert}{2c^2}\right)^4} \frac{k}{m}\z\left(\z+5\frac{K_3}{K_2}-\left(\frac{K_3}{K_2}\right)^2\z\right).
\een

We note that the heat flux in Fourier's law (\ref{53}) has the usual term proportional to the gradient of temperature and two relativistic terms, the first one proportional to an acceleration and the second one to the  gravitational potential gradient. The relativistic term proportional to the acceleration was proposed first by Eckart \cite{Eck} within an irreversible thermodynamic theory and is interpreted as an isothermal heat flux when matter is accelerated. The second relativistic term points out that in the absence of an acceleration, the heat flux in the presence of a gravitational field vanishes if the temperature gradient is counterbalanced by a gravitational potential gradient. This condition reduces to Tolman's law \cite{T1,T2}, namely, $$\frac{\nabla T}{T}=-\frac{\bf g}{c^2},$$ where $\bf g$ is the gravity field.

The coefficient of thermal conductivity (\ref{54}) has also a remarkable behavior in the presence of a gravitational field, because it decreases with the increase of the gravitational potential through the ratio $\vert\Phi\vert/c^2$. In the limiting case of weak gravitational potential $\vert\Phi\vert/c^2\ll1$ we have
\begin{enumerate}
  \item non-relativistic limit $\z\gg1$
  \ben
  \lambda=\frac{5kp\tau}{2m}\left[1+\frac{3}{2\z}+\dots\right]\left[1-\frac{2\vert\Phi\vert}{c^2} +\dots\right],
  \een
  \item ultra-relativistic limit $\z\ll1$
  \ben
  \lambda=\frac{4c^2p\tau}{T\z}\left[1-\frac{\z^2}{8}+\dots\right]\left[1-\frac{2\vert\Phi\vert}{c^2} +\dots\right].
  \een

\end{enumerate}

\subsection{Dependence of the temperature and particle number density  on the gravitational potential}
Let us consider a gas in the presence of the gravitational field in the absence of heat flux and acceleration field. Due to the spherical symmetry  the fields of temperature, gravitational potential and particle number density depend only on the radial coordinate, i.e., $T=T(r)$, $\Phi=\Phi(r)$ and  $n=n(r)$.

By taking into account the above considerations it follows from (\ref{53}) that the temperature field in a gravitational field must fulfill the differential equation
       \be
       \frac{1}{T(r)}\frac{dT(r)}{dr}=-\frac{1}{c^2}\frac{1}{1-\Phi(r)^2/4c^4}\frac{d\Phi(r)}{dr}.
       \ee{a}

The solution of the above differential equation for the boundary condition $T=T(R)$ and $\Phi=\Phi(R)$ at $r=R$ -- where  $R$ is the radius of the spherical source -- is
       \be
       \frac{T(r)}{T(R)}=\frac{\left(1-\frac{\vert\Phi(R)\vert}{2c^2}\right)\left(1+\frac{\vert\Phi(R)\vert}{2c^2}\frac{R}{r}\right)}{\left(1+\frac{\vert\Phi(R)\vert}{2c^2}\right)
       \left(1-\frac{\vert\Phi(R)\vert}{2c^2}\frac{R}{r}\right)},
       \ee{b}
which is a  decreasing function of $r$ for  $r>R$.

In a weak gravitational field where $\vert\Phi(R)\vert/c^2$ is a small quantity (\ref{b}) can be expressed as
       \be
       \frac{T(r)}{T(R)}=1-\frac{\vert\Phi(R)\vert}{c^2}\left(1-\frac{R}{r}\right)+\frac{\vert\Phi(R)^2\vert}{2c^2}\left(1-\frac{R}{r}\right)^2+\dots\,.
       \ee{c}

In general for stellar objects that are not too compacts the ratios $\vert\Phi(R)\vert/c^2$ evaluated at their surfaces are small as can be seen from Table 1.
\begin{table}[ht]
\centerline{\begin{tabular}{|c|c|c|c|} 
\hline
  & Radius & Mass &
$\vert\Phi(R)\vert/c^2$ \\
& (m) & (kg)& \\ \hline
Earth & $R_\oplus=6.38\times 10^6$ & $M_\oplus=5.97\times10^{24}$  & $7 \times 10^{-10}$ \\\hline
Sun & $R_{\odot}=6.96\times 10^8$ & $M_\odot=1.99\times10^{30}$ & $2.2 \times 10^{-6}$\\\hline
White dwarf & $5.4\times 10^6$&$1.02 M_{\odot}$   & $2.8 \times 10^{-4}$ \\\hline
Neutron  star &$2\times 10^4$& $M_{\odot}$ &  $7.5 \times 10^{-2}$\\ \hline
\end{tabular}}
\caption{Values for $\vert\Phi(R)\vert/c^2$ at surface of stellar objects.}
\end{table}

Let us analyze the balance momentum equation (\ref{48}) in order to estimate the dependence of the particle number density with the radial coordinate. Without the acceleration term, by taking into account (\ref{a}) and considering the equation of state $p=nkT$, it   becomes
\be
 \frac{1}{n(r)}\frac{dn(r)}{dr}=\left[\frac{mc^2}{kT(r)}\frac{K_3(mc^2/kT(r))}{K_2(mc^2/kT(r))}-1\right] \frac{1}{T(r)}\frac{dT(r)}{dr}.
\ee{d}

For a non-relativistic gas the ratio $K_3(mc^2/kT(r))/K_2(mc^2/kT(r))\approx 1$ and the solution of the differential equation (\ref{d}) reads
\be
\frac{n(r)}{n(R)}=\frac{T(r)}{T(R)}\exp\left[-\frac{mc^2}{kT(R)}\left(\frac{T(r)}{T(R)}-1\right)\right].
\ee{d1}
By considering a weak gravitational field $\vert\Phi(R)\vert/c^2\ll1$  and ${mc^2}/{kT(R)}\gg1$ eq. (\ref{d1}) reduces to
\be
\frac{n(r)}{n(R)}=1-\frac{m\vert\Phi(R)\vert}{kT(R)}\left(1-\frac{R}{r}\right)+\dots\,,
\ee{d2}
which is a decreasing function of $r$ for $r>R$. Note that this solution is valid only if the product $(\vert\Phi(R)\vert/c^2)({mc^2}/{kT(R)})\ll1$.

The solution of the differential equation (\ref{d}) for an ultra-relativistic gas is given by
\be
\frac{n(r)}{n(R)}=\left(\frac{T(r)}{T(R)}\right)^3,
\ee{d4}
since in this case the ratio $K_3(mc^2/kT(r))/K_2(mc^2/kT(r))\approx 4kT(r)/mc^2$. For a weak gravitational field $\vert\Phi(R)\vert/c^2\ll1$ the above equation can be approximated by
\be
 \frac{n(r)}{n(R)}=1-3\frac{\vert\Phi(R)\vert}{c^2}\left(1-\frac{R}{r}\right)+\dots\,,
\ee{d5}
showing a more accentuated decreasing with $r$ when compared with the one of the temperature field (\ref{c}).

\section{Conclusions}
In this work we derived constitutive equations for relativistic gases within the framework of Boltzmann equation. Two systems were analyzed. In the first one the Fourier and Ohm laws for
binary mixtures of electrons and protons and of electrons and photons subjected to external electromagnetic fields in special relativity were derived. Explicit expressions  for the thermal and electrical conductivities were given in the cases of binary mixtures of degenerate and non-degenerate electrons with non-degenerate protons and with photons. The other system consisted of a relativistic gas under the influence of a  spherically symmetrical non-rotating and uncharged source of the gravitational field described by the Schwarzschild metric. Here Fourier's law was derived and it was shown that the transport coefficient of thermal conductivity decreases in the presence of a gravitational field and that the heat flux
has three contributions, the usual dependence on the temperature gradient, and two relativistic contributions, one of them associated with an acceleration and another to a gravitational potential gradient. Furthermore, the dependence of the temperature field in the presence of a gravitational potential was discussed.

\section*{Acknowledgments}

This paper was partially supported by Conselho Nacional de Desenvolvimento Cient\'{\i}fico e Tecnol\'ogico (CNPq),  Brazil.

\end{document}